# ESO telbib: learning from experience, preparing for the future

Uta Grothkopf*[a], Silvia Meakins[a], Dominic Bordelon[a]

[a]European Southern Observatory (ESO), Karl-Schwarzschild-Str. 2, 85748 Garching near Munich, Germany

## ABSTRACT

The ESO telescope bibliography (telbib) dates back to 1996. During the 20+ years of its existence, it has undergone many changes. Most importantly, the telbib system has been enhanced to cater to new use cases and demands from its stakeholders. Based on achievements of the past, we will show how a system like telbib can not only stay relevant through the decades, but gain importance, and provide an essential tool for the observatory's management and the wider user community alike.

## 1. INTRODUCTION

The lifecycle of data at ESO starts with astronomers applying for observing time at the ESO telescopes. Once the proposal has been accepted and observing time granted, the observations are carried out, and the data are stored in the ESO Science Archive [1]. Observers typically enjoy a one-year proprietary time during which access to the data is restricted to the PI/CoIs. Subsequently, the observations are made available also to other researchers. Refereed papers that use partly or exclusively ESO data are identified by the curators of the ESO Telescope Bibliography (telbib) [2] and are added to the database. Links are established between the telbib records and the observing programmes in the Archive. Through a search at telbib, researchers retrieve information about articles that analyze data. Following the links to the Archive, they can request the data used by the authors for their own analysis. The reverse workflow is also possible: researchers query the ESO Science Archive for observations and are provided with information about the papers that deploy the data. In this way, proposals, data, and papers are interconnected. telbib is the only place where detailed information about all ESO data papers is available. The ESO Management regularly uses statistics derived from the telbib database for bibliometric studies in order to help evaluate the organization's scientific performance and impact.

In this paper, we give an overview of the ESO telescope bibliography and share our experience in setting up, enhancing, and maintaining this complex system. In particular, we emphasize why it is important that bibliographies be curated, explain the various roles the curators fulfill, and describe the lessons we learned during more than 20 years of telbib's existence. We dubbed these lessons "The 7 Cs". Finally, we draw some conclusions.

## 2. THE ESO TELESCOPE BIBLIOGRAPHY IN A NUTSHELL

The ESO Librarians develop and curate the ESO telescope bibliography database. As of May 2018, telbib includes metadata of over 14,000 refereed papers published since 1996. In 2017, for the first time in the history of the organization, the ESO users community deployed ESO data in more than 1,000 papers in a single year. New telbib records are added based on text mining, followed by thorough inspection by the librarians who make sure that records are linked to all observational data that were used in the paper, and also eliminate "false hits" that the text mining may have erroneously identified. In a recent paper, Patat et al. [3] found that telbib's completeness is better than 96%. With the currently available automated tools, such a high fraction, both regarding the number of retrieved papers and the correctness of links to data, cannot be achieved.

*uta.grothkopf@eso.org; phone +49 89 32006-280; www.eso.org/libraries

Through the large number of parameters that are available in the database, telbib provides various reports and statistics that help to understand the performance and impact of ESO's scientific output through bibliometric studies. More information about telbib, including the paper classification policy, can be found on the web [4].

## 3. ROLES IN CONTENT CREATION

To better understand the contributions and roles related to content creation and curation, we adapt a graph presented in an article by Julie McMurry et al. [5] in a recent *PLoS Biology* article. In Fig. 3 of this paper, the authors identify eight roles that are essential in the process of creating and providing content [6]. These roles can be related to various entities and tasks in the publishing process.

The Author (role no. 1) provides the original content which is then handed over to the Guardian (no. 2, the publisher) to take care of it and process it, and on to the Curator (no. 3, libraries and/or repositories). Roles no. 4-8 fall in the responsibility of the librarians in charge of telbib: the Annotator (4) expands the content (telbib: tags are assigned to the records to optimize retrieval), and the Integrator (5) combines it with other existing content in novel ways (telbib: papers are linked to data and other information resources in order to create a unique place of interlinked resources). The Contributor (6) makes factual corrections and further improvements (telbib: wrong programme identification codes [see Note 7] are corrected and missing ones added). The content is then aggregated and prepared for indexing and searching by the Indexer (7) (telbib: records are made searchable based on a multitude of parameters), and finally the Application Provider (8) creates a platform for public access and a unified user experience (telbib: database records are retrievable via the telbib.eso.org platform). All of these roles add indispensable value and are strong arguments for humanly curated telescope bibliographies.

## 4. LESSONS LEARNED: "THE 7 CS"

In over 20 years of telbib's existence, the curators had ample opportunity to gain experience and expertise regarding creating, improving, and deploying a telescope bibliography. In this section, we describe the lessons we learned. These are dubbed "The 7 Cs".

### 4.1 Lesson no. 1: Connectivity

One of the main requirements at the time of creation of the ESO telescope bibliography was the need for interoperability between telbib and the ESO Science Archive in order to connect telbib records seamlessly with the data in the Archive. The database system underlying the Archive is a Sybase relational database, hence telbib was built in the same database environment.

In the following years, further links were established between telbib and other information resources. In all cases, easy data exchange is crucial. Bibliographic reference codes (bibcodes) are used to exchange information about telbib records with the NASA Astrophysics Data System (ADS) [8] and within ESO to link to ESO Press Releases [9], ORCID author IDs [10] link to author profiles, and Digital Object Identifiers (DOIs) connect telbib to the network of scholarly papers [11] and Altmetric article metrics [12]. In addition, APIs (Application Programming Interfaces) are used to link to and from telbib. Fig. 1 provides an overview of the many links to and from telbib records.

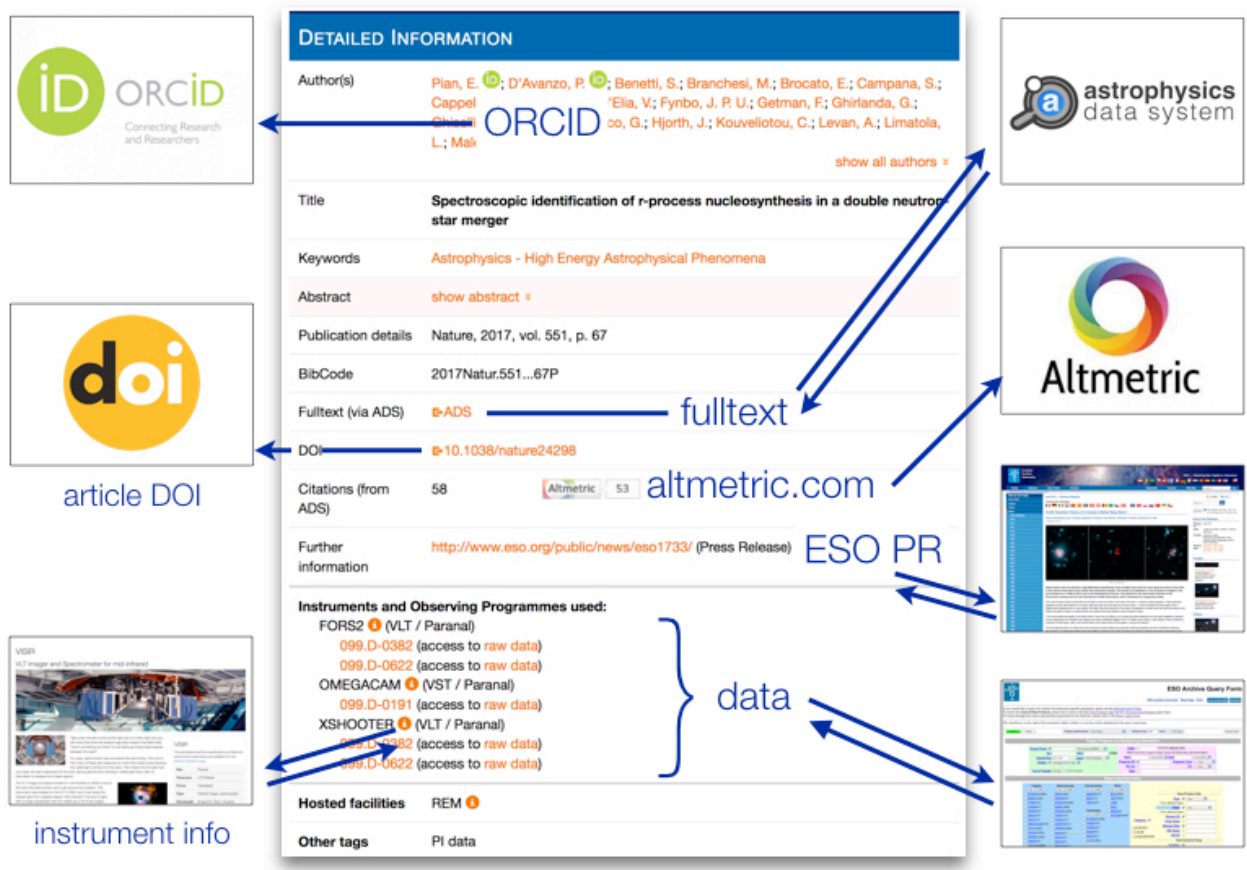

Figure 1. telbib sample record, showing the interconnectedness of records (counterclockwise, starting in the lower right corner): data links to and from the ESO Science Archive, links to and from ESO Press Releases, social media attention measurements from Altmetric, article fulltexts from the ADS and links from the ADS to telbib for data access, links to author profiles at ORCID, Digital Object Identifiers (DOIs) for persistent identification and access, and links to further information about the instruments used in the paper.

## 4.2 Lesson no. 2: Curation

In today's world of text mining, automated retrieval and artificial intelligence systems applied to big data, it might seem that human intervention and verification during the record creation process is an antiquated, if not obsolete, concept. The text mining tool developed and maintained by the ESO Library for use at ESO, called FUSE, indeed is extremely helpful, but it merely points the telbib curators to those articles that need to be inspected regarding the possible inclusion into telbib. FUSE certainly does not replace the human vetting which is essential in order to assure a correct and complete set of data papers in the bibliography – in particular because no text mining software can find identifiers and other information that the authors did not put into their manuscripts in the first place.

telbib records receive annotations, indicating in which way authors referenced ESO data. Such references range from correctly mentioning in a footnote or in the acknowledgements (as is required by the ESO data citation policy [13]) all programme IDs that provided observations, to merely mentioning the observatory or the ESO Science Archive in the text of the paper. A brief study done by the ESO Librarians revealed that during the publishing years 2012 to 2017, only approx. 70% of the VLT/VLTI and VISTA/VST papers reference the correct and complete programme identifiers in the paper. In other words, approx. 30% of the data papers would not be linked properly to the respective data in the Archive through automated text mining and linking. In these cases, the telbib curators retrieve programme IDs (e.g., through Archive searches based on observational details provided by the authors) even though they were not mentioned in the papers. For ALMA, the fraction of papers with correct and complete references of programme identifiers is considerably higher, namely approx. 91%. It can be safely assumed that the reason for this striking difference lies in the strictness with which data citation policies are implemented by the respective observatories. Since the beginning of data paper publication, the ALMA Management has made large efforts to establish proper data citation. In contrast, ESO has made data citation a requirement only a few years ago, while before acknowledging ESO data was less obligatory. For APEX, the fraction of papers that correctly and completely cite observations in publications is considerably lower (around 33% according to our study), indicating a heterogeneous way of citing data, and a rather reluctant enforcement of the citation policy by the various APEX partner organizations (Fig. 2).

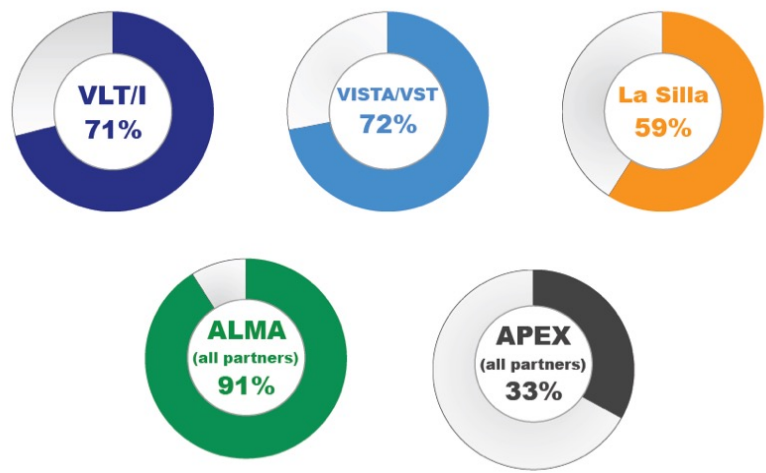

Figure 2. Fraction of VLT/VLTI, VISTA/VST, La Silla, ALMA, and APEX data papers with correct and complete programme IDs (publication years 2012-2017). For ALMA and APEX, papers using data from all partner organizations are taken into account.

### 4.3 Lesson no. 3: Community Involvement

"Author awareness" and "dissemination of data citation policies" are keywords that describe Lesson no. 3. They are closely related to Lesson no. 2 -- in order for authors to properly cite the data they used, obviously they need to be aware of the existence of an observatory's data citation policy.

At ESO, many communication channels are used to disseminate the information about how observations should be cited in papers, including a dedicated web page [13], instrument scientists who inform their users community, and the ESO Observing Proposals Office and Science Archive that alert astronomers regarding the official ESO acknowledgement at various stages of observation preparation and data acquisition. We also very much appreciate the efforts of some "power users" among the ESO users community who have been observing with ESO's facilities frequently and for many years; many of them are well aware of the data citation policy and often "educate" their collaborators and students accordingly.

In addition, the telbib curators often communicate directly with authors, in particular in order to clarify questions regarding the use of data in specific papers. Such email conversation is an excellent opportunity to remind authors about the required acknowledgement. As an example, the ESO Library investigated records added to telbib between September and December 2017. The total number of new additions was 414. In 45 cases (11%), the telbib curators contacted the authors to enquire about the observations that were used, and to alert them about the ESO data citation policy. By March 2018, we had received 28 replies, or 62%. In other words, in about two thirds of the cases authors were reminded about correct data acknowledgement, and according to our experience, in most cases were very eager to fulfil ESO's requirements, and properly acknowledge data use in future papers.

## 4.4 Lesson no. 4: Collaboration

The concept of collaboration is closely associated with teamwork, exchange of best practices, and networking. In many areas in life, it is essential to join efforts in order to achieve the best results, and telescope bibliographies are not any different. For telbib, we often communicate with other key players at ESO and beyond to establish new features, tags, and metadata. We also cooperate with bibliography curators at other observatories to define policies and guidelines for our respective bibliographies.

An excellent example for such collaboration is the ALMA bibliography which is jointly maintained by the librarians at ESO and NRAO, along with NAOJ. Issues arising from incomplete or erroneous data citation in papers are discussed and solved jointly. Even more, the different routines and workflows at the three organizations complement each other and lead to a particularly thorough workflow.

## 4.5 Lesson no. 5: Conjecture (Anticipation)

Lesson no. 5 is called "Conjecture" for the sake of maintaining the set of "Cs", but it could also have been named "Anticipation". In order to ensure that a database like telbib continues to be relevant through time, it needs to be enhanced with new features, according to the questions the ESO Management and other stakeholders need to answer. Often such questions arise on short notice, i.e., at the time when statistics are already needed. It is therefore absolutely essential for bibliography curators to follow developments in science and in the publishing sector not only at their own organisation, but in astronomy in general. Ideally, new metrics can then be developed even before the demand is expressed by the requesters.

Based on this approach, telbib has grown over time from a list of articles to a sophisticated system. New features are often implemented proactively and are deployed in many ways. Use cases include for instance statistics compiled *ad-hoc* based on user requests, regularly published reports, or articles that use statistics derived from telbib, such as the investigation of the impact of science operations on science return at the VLT by Sterzik et al. in 2016 [14], and the 2017 report by Patat et al. on non-publishing programmes [3].

## 4.6 Lesson no. 6: Courage

As the telbib curators, we are always happy to see that the database is valued highly and used frequently, and that it provides meaningful information to the requesters. Obtaining (positive) feedback is equally rewarding. On the other hand, sometimes telbib users provide suggestions for improvements which we would like to implement, if possible. However, it is obvious that we are facing an ever increasing number of published papers that need to be inspected, along with a rising number of parameters that have to be checked for each added paper.

Therefore, not all suggested modifications can actually be implemented, and it is essential for the bibliography curators to find the courage to decline requests if they are not feasible, or if their implementation would absorb time and effort on the side of staff that would likely not be balanced by the assumed benefits. In these cases, curators need to learn to say 'no'.

## 4.7 Lesson no. 7: Count on Your Team

Our final lesson is: it is essential to count on your team. The ESO Library Garching has a team of three librarians, each of them with their own skills, knowledge, and professional role. While we are all specialized in specific tasks, some of our knowledge overlaps with that of our colleagues. This redundancy is necessary to be able to take over the others' tasks if necessary, and to avoid a single point of failure.

# CONCLUSIONS

In this paper, we have shown that the ESO telescope bibliography (telbib) is a curated database. The concept of curation is interpreted as "to take care of", and indeed the telbib curators take this task very seriously. The telbib workflow applied at the ESO Library ensures quality content which includes (but is not limited to) links between the published articles and the corresponding observational data in the ESO Science Archive.

During more than 20 years of maintaining the ESO telescope bibliography, and developing it from a rather simple list of papers to a sophisticated database system, we have gained much experience regarding the key ingredients that form the basis of a successful bibliography. We have dubbed these insights "The 7 Cs": (1) Connectivity; (2) Curation; (3) Community involvement; (4) Collaboration; (5) Conjecture (Anticipation); (6) Courage, and (7) Count on Your Team.

To fulfill these requirements, a wide range of skills are necessary, some of them pertaining to the technical and programming sector (in particular for the areas 1 and 5), others demand experience and knowledge of the scientific community (specifically areas 2 and 5) as well as "soft skills" such as empathy and communication skills (areas 3, 4, 6). All skills are combined in area 7, the team spirit among those in charge of curating the telescope bibliography.

As we are moving the ESO telescope bibliography towards the completion of the third decade of its existence, we will continue to proactively implement new features. Whenever possible, we will positively react to new or changed demand from telbib stakeholders, but we will also maintain the liberty to select only requests that are feasible and that will add long-term value to the database.

# ACKNOWLEDGEMENTS

The ESO telbib curators would like to cordially thank the many colleagues at ESO and beyond who helped to shape the database into a versatile tool. In particular, our thanks go to Angelika Treumann who was in charge of telbib during its early years and paved the way for it to become an important tool, as well as to Chris Erdmann for developing and building the first version of the complex system telbib is today.